\documentclass[printer]{aa}
\usepackage{amsmath}
\usepackage{graphicx}
\usepackage{natbib}
\usepackage{txfonts}
\usepackage{default}
\usepackage{acronym}
\usepackage{aas_macros}
\usepackage{xspace}

\newcommand{\kev}[1][]{#1\,\ensuremath{\mathrm{keV}}}
\newcommand{\unit}[2][]{#1\,\ensuremath{\mathrm{#2}}}
\newcommand{\pten}[2]{\ensuremath{\mathrm{#1 \times 10^{#2}}}}

\newcommand{\amin}[1][]{#1\arcmin}
\newcommand{\asec}[1][]{#1\arcsec}

\newcommand{\tnote}[1]{\ensuremath{^{#1}}}
\newcommand{\tnotetxt}[2]{\footnotesize{\ensuremath{^{#1}} #2}\\}
\newcommand{\src}[1]{\acl{#1}}
\newcommand{\srcf}[1]{\acs{#1}}

\newcommand{\XMM}{XMM-\textit{Newton}\xspace}
\newcommand{\Chandra}{\textit{Chandra}\xspace}
\newcommand{\pn}{EPIC-pn\xspace}
\newcommand{\mos}{EPIC-MOS\xspace}

\acrodef{WFC}{wide field camera}
\acrodef{EPIC}{European photon imaging camera}
\acrodef{SAS}{the \XMM science analysis software}
\acrodef{ARF}{ancillary response file}
\acrodef{RMF}{response matrix file}
\acrodef{DEM}{differential emission measure}
\acrodef{CIE}{collisionally ionised equilibrium}
\acrodef{ICM}{intra-cluster medium}
\acrodef{AGN}{active galactic nuclei}
\acrodef{HRI}{high resolution instrument}
\acrodef{PSF}{point spread function}

\acrodef{0301}[\object{1RXS\,J081232.3$-$571423}]{J0812}
\acrodef{0501}[\object{1RXS\,J160147.6$-$754507}]{J1601}
\acrodef{1101}[\object{1RXS\,J153934.7$-$833535}]{J1539}

  \abstract{} {We studied the physical properties of three clusters of galaxies, selected from a BeppoSAX \ac{WFC} survey. These sources are identified as \srcf{1101}, \srcf{0501}, and \srcf{0301} in the ROSAT All-Sky Survey catalogue. We obtained \XMM follow-up observations for these three clusters.} {We fitted single and multi-temperature models to spectra obtained with the \pn camera to determine the temperature, the chemical composition of the gas and their radial distribution. Since two observations were contaminated by a high soft-proton background, we developed a new method to estimate the effect of this background on the data. }{We present the temperature and iron abundance of two of these three clusters for the first time. The iron abundance of \srcf{1101} decreases with radius. The fits to the \XMM and \Chandra data show that the radial temperature profile within \amin[3] towards the centre either flattens or lowers. A \Chandra image of the source suggests the presence of X-ray cavities. The gas properties in \srcf{0501} are consistent with a flat radial distribution of iron and temperature within \amin[2] from the centre. \srcf{0301} is a relatively cool cluster with a temperature of about \kev[3].} {The radial temperature and iron profiles suggest that \srcf{1101} is a cool core cluster. The \Chandra image shows substructure which points toward \ac{AGN} feedback in the core. The flat radial profiles of the temperature and iron abundance in \srcf{0501} are similar to the profiles of non-cool-core clusters. \acresetall}


\title {X-ray spectral study of hot gas in three clusters of galaxies}
\author{Y.~G. Grange \inst{1} \and E. Costantini \inst{1} \and J. de Plaa \inst{1} \and J.~J.~M. in 't Zand \inst{1}  \and F. Verbunt \inst{2} \and J.~S. Kaastra \inst{1,2} \and F. Verrecchia \inst{3}}
\institute{SRON Netherlands Institute for Space Research, Sorbonnelaan 2, 3584 CA Utrecht, The Netherlands 
\and Astronomical Institute, Utrecht University, PO Box 80000, 3508 TA Utrecht, The Netherlands
\and ASI Science Data Center, ESRIN, I-00044 Frascati(RM), Italy
}
   \date{Received 4 September 2009 / Accepted 12 January 2010}

\begin{document}
\maketitle
\section{Introduction} \label{sec:int}
Clusters of galaxies are the largest gravitationally bound objects in the universe. They are the result of the growth of over-densities that were present in the early universe. About 80\% of the mass in clusters consists of dark matter. Until now it has been impossible to directly measure it. The baryonic matter in clusters of galaxies resides in the stars and a hot diffuse plasma (hereafter referred to as gas). The stars account for only 1--3\% of the total cluster mass, while the gas is responsible for 10--15\% \citep[e.g.][]{DAV90,ALL02}. This gas, called the \ac{ICM}, has temperatures of $\sim 10^{7-8}$K, high enough to emit X-ray radiation. 
\par Originally, clusters of galaxies were discovered based on their optical properties \citep[e.g.][]{ABE58}. In the 1960s and 1970s, X-ray emission was detected from positions or directions coinciding with those of known clusters of galaxies \citep{BYR66,BRA67,KEL71}. \citet{CAV71} stated that this emission was indeed originating from the clusters. In the past ten years, the instruments on board of the \XMM and the \Chandra X-ray telescopes enabled the study of the physical and chemical properties of the gas in more detail.
\par In the X-ray band, two general catalogues have been compiled. In the hard X-ray band (0.5 to 25 \kev[]), HEAO-1 surveyed the sky with a detection threshold of \unit[\pten{5}{-15}]{W\,m^{-2}} \citep{WOO84}. The soft X-ray band (\kev[0.1 -- 2.4]) was probed in the ROSAT All Sky Survey, with a flux limit of \unit[\pten{5}{-16}]{W\,m^{-2}} \citep{VOG99}. The ROSAT observations yielded various catalogues of clusters of galaxies \citep[e.g.][]{REI02,EBE02,BOH04}. 
\par We present the results of a spectral analysis of three clusters of galaxies, selected from observations with the \acfp{WFC} on board BeppoSAX. The Italian-Dutch X-ray observatory BeppoSAX \citep{BOE97} operated between June 1996 and April 2002 and carried two \acp{WFC} \citep{JAG97}. They had a field of view of $40 \degr\times40  \degr$ (full width at zero response), angular resolution of 5\arcmin\ (FWHM) and pointed in opposite directions. They were sensitive in a bandpass of 2 to 28 keV with a spectral resolution of \kev[1.2] at 6 keV (FWHM). The \acp{WFC} observed the whole sky. The exposure coverage was somewhat inhomogeneous. However, most positions on the sky have been observed for at least 1 million seconds. The sensitivity level reaches \unit[\pten{3}{-15}]{W\,m^{-2}}. Therefore only the brightest sources of ROSAT are detected by the \ac{WFC}.
\par All data of this instrument were combined on time scales of six months as well as the whole mission and in energy ranges of 2--5, 5--9 and \kev[9--28]. The combined data yielded detections of a few of cataclysmic variables and tens of \ac{AGN} and clusters of galaxies \citep[see e.g.][for a catalogue based on part of this survey]{VER07}. Six objects were found to be ill-understood despite their brightness. The \kev[2--10] fluxes of these six sources lie in the \pten{7}{-15} to \unit[\pten{3}{-14}]{W\,m^{-2}} range. 
\par We followed up these bright sources using \XMM. For one source we analysed the available \Chandra archival data. Three of these six are clusters of galaxies. The physical properties of the gas in these clusters were found using the X-ray spectra.
\par Throughout this paper, we use $H_{0}=\unit[70]{km\,s^{-1}\,Mpc^{-1}}$, $\Omega_{\mathrm{M}}=0.3$, and $\Omega_{\Lambda}=0.7$. The abundances are in units of the proto-solar abundances by \citet{LOD03}. Unless specified otherwise, quoted statistical uncertainties are at a $1\sigma$ (68\%)  confidence level.
\section{The clusters discussed in this paper}
Here, we introduce the three observed clusters and give an overview of the catalogues in which they appear. We refer to the three clusters by the first four numerals from their 1RXS-catalogue name preceded by the letter ``J''. The fluxes and luminosities found by previous authors are quoted in Table \ref{tab:wfc}. 
\subsection{\src{1101} (\srcf{1101})}
This source is part of the REFLEX catalogue \citep[\object{RXC\,J1539.5$-$8335}, ][]{BOH04}, a cluster survey based on ROSAT data.
\subsection{\src{0501} (\srcf{0501})}
\begin{figure}[t]
 \centerline{\includegraphics[width=0.9\linewidth]{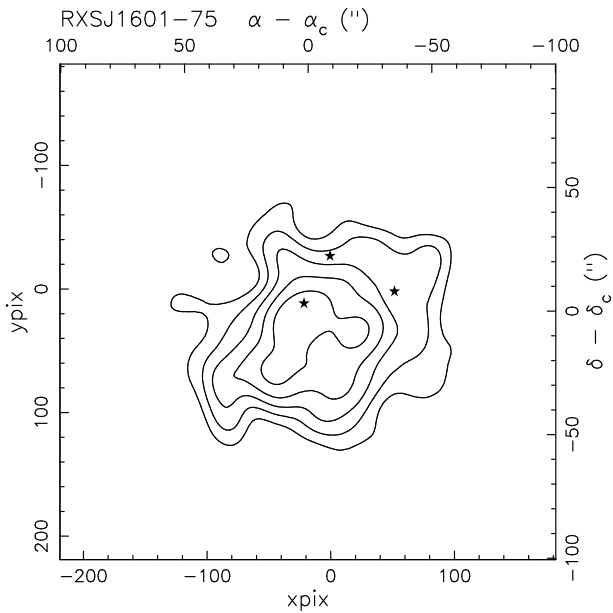}}
\caption{Contours of an archival ROSAT \ac{HRI}  observation of \src{0501}. The three dots plotted are optically bright foreground stars. The observation is binned  using a Gaussian with a \asec[6] standard deviation. The $n$th contour corresponds to $0.8^{n}$ times the maximal count rate.} \label{fig:0501_rosat}
\end{figure}
\src{0501} is part of the ROSAT catalogue of clusters in the ``zone of avoidance'', meaning a location within 20\degr{} of the Galactic plane \citep[\object{CIZA\,J1601.7$-$7544},][]{EBE02}. That is also the first detection and identification of this source as a cluster of galaxies. The source also appears in the RXTE All-Sky Slew Survey catalogue, \citep[\object{XSS\,J16019$-$7548},][]{REV04}. Figure \ref{fig:0501_rosat} shows intensity contours of an archival observation of this source, using the \ac{HRI} aboard ROSAT.
\subsection{\src{0301} (\srcf{0301})}
\src{0301} was first detected by the Einstein X-ray satellite as part of the Einstein Galactic plane survey \citep[\object{1E\,0811.5$-$5704},][]{HER84}. Furthermore, it appears in the Einstein Slew Survey \citep[\object{1ES\,0811.4$-$5704},][]{ELV92} and is also part of the 2E catalogue \citep[\object{2E\,0811.4$-$5704},][]{HAR90}. The source was identified as a cluster of galaxies by  \citet{EBE02} and incorporated in the CIZA catalogue (\object{CIZA\,J0812.5$-$5714}).
\begin{table}[t]
\begin{center}
\caption{Previously known parameters for the studied sources.} \label{tab:wfc}
\begin{tabular}{lllll}
\hline
Object  & WFC Flux \tnote{a}  & ROSAT flux\tnote{b} & redshift \tnote{c} \\
\hline
\hline
\src{1101} & 0.82 & 1.68 \tnote{d}$^{,}$\tnote{e} & 0.0728 \tnote{d} \\
& &  2.49 \tnote{f} &  \\
\src{0501} & 1.28 &  2.41 \tnote{f} &  0.1530 \tnote{g} \\
\src{0301} & 1.14 & 3.93 \tnote{f} & 0.0620 \tnote{g} \\
\hline
\end{tabular} \\
\end{center}
\tnotetxt{a}{\kev[2--10], 10$^{-14}$ W\,m$^{-2}$ for Crab-like spectrum (i.e. a power law with a photon index of 2)}
\tnotetxt{b}{\kev[0.1--2.4], 10$^{-14}$ W\,m$^{-2}$}
\tnotetxt{c}{Determined using optical spectroscopy}
\tnotetxt{d}{From REFLEX \citep{BOH04}}
\tnotetxt{e}{Using a Raymond-Smith model \citep{RAY77}}
\tnotetxt{f}{From the RASS-BSC \citep{VOG99}, assuming a power-law spectrum.}
\tnotetxt{g}{From CIZA \citep{EBE02}}
\end{table}
\section{Data reduction and analysis}\label{sec:data}
\subsection{\XMM \acs{EPIC}}\label{subsec:EPIC} 
A log of the \XMM \acl{EPIC} \citep[\acs{EPIC};][]{STR01} observations is given in Table \ref{tab:pndata}. For all our observations the ``medium'' filter was put in place. For each observed source, we extract a lightcurve to look for soft-proton flares. Since emission from clusters of galaxies is not time dependent, we use the full energy range of the detector to extract this light curve. We compared the lightcurves within \amin[3] around the cluster centre to the lightcurve of the remainder of the field. Both light curves show flaring at the same time scales which excludes the possibility of a variable point source affecting the light curve. For two observations (\src{1101} and \src{0501}), the residual soft-proton background was very high (see section \ref{subsec:flares}). We excluded data during the flares on top of this high background. The original and filtered observation times are listed in Table \ref{tab:pndata}.
\begin{table}[t]
\begin{center}
\caption{Exposures of the XMM follow-up of the \ac{WFC} data.} \label{tab:pndata}
 \begin{tabular}{llll}
\hline
 Object & time (ks) \tnote{a}  & filtered time (ks) & date (Y-M-D)\\
\hline
\hline
\src{1101} & 17.7 & 13.8 & 2008-04-01\\
\src{0501}\tnote{b} & 9.12  &  6.96 & 2008-02-21\\
\src{0301} & 1.53 & 1.53 & 2008-03-01\\ 
\hline
\end{tabular}
\end{center}
 \tnotetxt{a}{Sum of good time intervals for central CCD.}
 \tnotetxt{b}{In our observation, the pointing was towards another unidentified ROSAT source (\object{1RXS\,J160107.6$-$754534}) within the uncertainty circle of the \ac{WFC}.}
\end{table}
\subsubsection{Standard data treatment }\label{subsec:reduction}
The data reduction is done with version 8.0.1 of \ac{SAS}. We used the calibration version of April 1st of 2009 and the standard \pn pipeline to create event files. During the observations, \mos was in small window mode. We did not use these data because in this mode the source is larger than the field of view. We removed the point sources by eye using a circle with a \asec[15] radius around them. The next step involved correcting for the vignetting of the photons and protons. For the observation of \src{0301}, which is not dominated by a high soft-proton background, we used the \texttt{evigweight} task to take this effect into account. This task assigns a theoretical vignetting factor to each photon based on its energy and position on the CCD.  For the other two observations, we took care of the soft-proton vignetting as described in Sect. \ref{subsec:flares}. 
We selected only the single and double events (\texttt{PATTERN$\leq$4}) that were inside the field of view (\texttt{FLAG==0}).
For each separate spectrum, we generated an \ac{ARF} and a \ac{RMF} using a detector map. For \src{0301}, we took the detector map to be flat (i.e. without any spatial distribution) because spatial effects are already taken care of by the vignetting correction of \texttt{evigweight}.
For \src{1101} and \src{0501}, we made an image of the event file and used that as a detector map. 
\par For the spectral reduction, we rebinned the spectra, sampling the instrument FWHM by a factor of three, requiring at least 20 counts per bin to ensure the applicability of $\chi^{2}$ statistics if needed, and checked that spectral features were not undersampled.
\par The next step was the background treatment. Since the sources are relatively compact, we took the background from an annulus ranging from \amin[9] to \amin[12], centred on the telescope pointing. This is the outer edge of the field of view where we did not expect any residual cluster emission. We scaled the background spectrum with the ratio of the number of illuminated pixels  in the observations to the background annulus. We subtracted the scaled background spectrum from the observation. Since there is a strong instrumental copper line at \kev[8.04] in the background region, we ignored the energy range from \kev[7.8] to \kev[8.4] for analysis of the spectra to ensure that no radiation of this line was still present. For other instrumental lines, this effect is expected to be negligible in the energy range of the fit due to the statistics of the observations. 
\subsubsection{Special treatment of flared events} \label{subsec:flares}
Normally the  intervals of soft-proton flaring in the light curve are excluded from the analysis. The observations of \src{1101} and \src{0501} have a very high background. Inspection of the light curves of these observations for the region outside the \amin[3] around the emission peak of the observation yields an average count rate of \unit[$\sim$6]{c\,s^{-1}}. In the \ac{SAS} User Guide, the rule of thumb is to ignore all data for which the count rate is higher than \unit[1]{c\,s^{{-1}}}. This would result in discarding all the data despite a positive detection of the source. In order to be able to use the data, we developed a method to estimate this background contribution.
\par \citet{KUN08} studied the soft protons in \XMM data. They found a vignetting effect for soft protons in the \mos detector. For the \pn, such a study has not yet been performed. Since the vignetting is caused by focussing of the soft-proton events by the mirror, we expect a vignetting similar to \mos. 
Protons are vignetted less than X-ray photons. Therefore, standard \ac{SAS} tasks for correcting the X-ray photon vignetting are not applicable. To be able to fit the spectra, we derived a vignetting factor for the soft-proton background of these data sets. We assumed that the X-ray photon background is negligible compared to the soft-proton background. 
Since the average background count rate is of a factor six higher than the maximum expected X-ray quiescent level, this is a reasonable assumption. Other simplifications are that the vignetting of soft-protons is taken to be energy-independent and 
that the vignetting in the innermost \amin[3] around the cluster centre is negligible. The radial behaviour of the soft-proton vignetting measured for MOS by \citet{KUN08} is consistent with being flat within the inner \amin[3]. 
\begin{figure}[t]
\centerline{\includegraphics[angle=-90, width=\columnwidth]{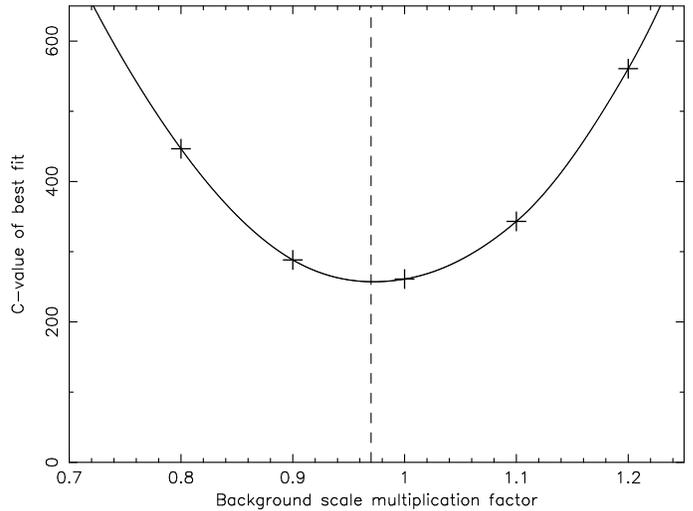}}
\caption{Example of the Cash-statistic minimisation method, as used for cluster \src{1101}. The multiplication factors are based on the region \amin[3] around the cluster centre.} \label{fig:minim}
\end{figure}
\par In order to find the vignetting factor, we first scaled the background to the region around \amin[3] from the cluster centre, only based on the number of illuminated pixels. Then we created a set of spectra where we subtracted this background multiplied by a range of factors and fitted a model (as described in Sect. \ref{sec:models}) to all of them. The C-statistic values of the fits as a function of the scale factor for \src{1101} are shown in Fig. \ref{fig:minim}. We fitted a parabola to the data points to find the vignetting factor and the uncertainty on this value. The multiplication factor at the minimum of the parabola was taken as the best-fit value. The $1\sigma$ uncertainty is then given by the scale factor value for which the C-statistic is 1 above the minimal value \citep{CAS79}. 
\begin{figure}[t]
 \centerline{\includegraphics[width=\columnwidth]{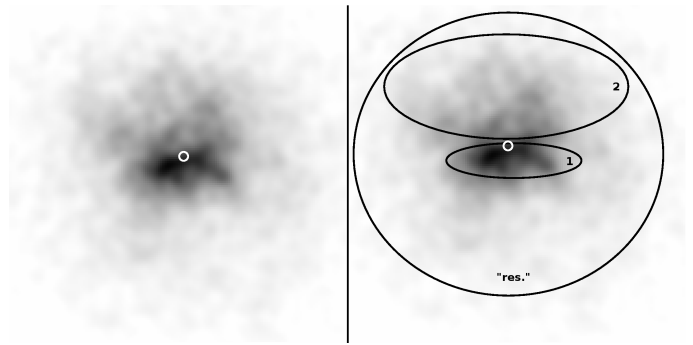}}
\caption{\Chandra image of \src{1101}, based on a \unit[8]{ks} observation using the ACIS instrument. There are no X-ray point sources in the region shown. The image is smoothed using a Gaussian with a standard deviation of \unit[2.5]{pixels}. The white circle represents the position of the central cD galaxy in this cluster. In the left panel, a hint of sub-structure is visible. The regions we used to fit the different parts of the spectrum are shown in the right panel. The projected radii of the elliptical ``res.'' region are \unit[97]{kpc} (width) and \unit[89]{kpc} (height). }\label{fig:chandra}
\end{figure}
\subsection{\Chandra}
\Chandra data are available only for \src{1101}. The \unit[8]{ks} ACIS observation was taken in pointing mode on 24 June 2007.  The \Chandra observation does not suffer from high background. This together with a high spatial resolution allowed us to investigate the spatial properties of the object. We then performed the spectral analysis of the regions marked in Fig. \ref{fig:chandra}. To compare both data sets, we also analysed the spectra based on the same annuli as used in the \XMM analysis. This comparison is discussed in Sect. \ref{sec:res_1539}. 
\subsection{Spectral models }\label{sec:models}
We fitted the spectra with the SPEX package \citep{KAA96}. We employ two different models. The first is the \ac{CIE} plasma model based on the MEKAL code \citep{MEW95}, implemented as the \textit{cie} model in SPEX. This model represents an isothermal gas, which is appropriate for low-statistics data, where we cannot distinguish between single and multi-temperature models. For data with higher statistics, a \ac{DEM} type model can be used \citep[e.g.][]{KAA04}. In this model the emission measure $Y=\int n_{\mathrm{e}} n_{\mathrm{H}} \mathrm{d}V$ (where $n_{\mathrm{e}}$ and $n_{\mathrm{H}}$ are the electron and proton densities and $V$ is the source volume) is a function of temperature. The \textit{wdem} model \citep{KAA04} is a parametrisation of this type, which provides a good empirical description of the cores of clusters of galaxies \citep[e.g.][]{PLA04,KAA04}. The model consists of thermal components distributed as a truncated power law. The model is characterised by
\begin{equation}
\frac{\mathrm{d}Y}{\mathrm{d}T} = \left\{ \begin{array}{ll} 
cT^{1/\alpha} & \hspace{1.0cm} \beta T_{\mathrm{max}} \le T < T_{\mathrm{max}} \\
0 & \hspace{1.0cm} T > T_{\mathrm{max}} \lor T < \beta T_{\mathrm{max}} \\
\end{array} \right..
\label{eq:wdem_dydt}
\end{equation}
The lower temperature cutoff of the distribution is located at $\beta kT_{\mathrm{max}}$. This form of the model is chosen in a way that if $\alpha$ approaches 0, the model becomes isothermal with temperature $kT_{\mathrm{max}}$. For this study, we empirically found that $\beta=0.18$ yields a model which has negligible degeneracy between the maximum temperature ($kT_{\mathrm{max}}$) and the power law slope ($1/\alpha$). This value agrees with earlier work \citep{PLA06,WER06}.
\par For a comparison between the outcomes of the (multi-temperature) \textit{wdem} model to single temperature-models, we employed the mean temperature, $T_{\mathrm{mean}}$. This value is defined as
\begin{equation}
T_{\mathrm{mean}} = \frac{\int T \frac{\mathrm{d}Y}{\mathrm{d}T} \mathrm{d}T}{\int \frac{\mathrm{d}Y}{\mathrm{d}T} \mathrm{d}T}.
\label{eq:em_tmean_int}
\end{equation}
The evaluation of the integral between $\beta kT_{\mathrm{max}}$ and $kT_{\mathrm{max}}$ yields, in terms of the model parameters
\begin{equation}
kT_{\mathrm{mean}} = \frac{(1 + 1/\alpha)}{(2 + 1/\alpha)} \frac{(1 - \beta^{1/\alpha + 2})}{(1 - \beta^{1/\alpha + 1})} kT_{\mathrm{max}}.
\label{eq:wdem_mean}
\end{equation}
\par We used a neutral absorbing gas model for the determination of the interstellar hydrogen column density. Since the hydrogen column density value from \ion{H}{i} maps \citep[e.g.][]{DIC90,KAL05} does not include molecular hydrogen, it is possibly lower than the value needed to fit the spectrum. Therefore we kept $N_{\mathrm{H}}$ as a free parameter in our fits.
\section{Results } \label{sec:res}
\begin{figure}[ht]
 \centerline{\includegraphics[angle=-90,width=0.9\linewidth]{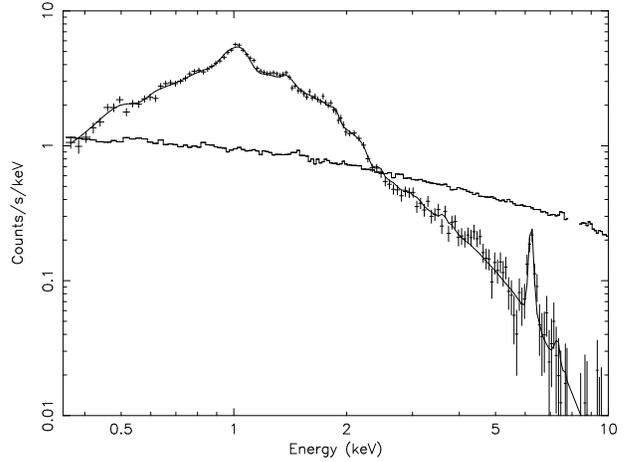}}
\caption{The \pn spectrum of \src{1101} (crosses) with subtracted background level (histogram) as explained in Sect. \ref{subsec:flares} and the fitted model (connected line). The part above \kev[2] is rebinned with a factor two for presentational purposes.} \label{fig:spec_1101}
\end{figure}
\begin{figure*}[ht]
\centerline{
 \includegraphics[width=0.3\linewidth]{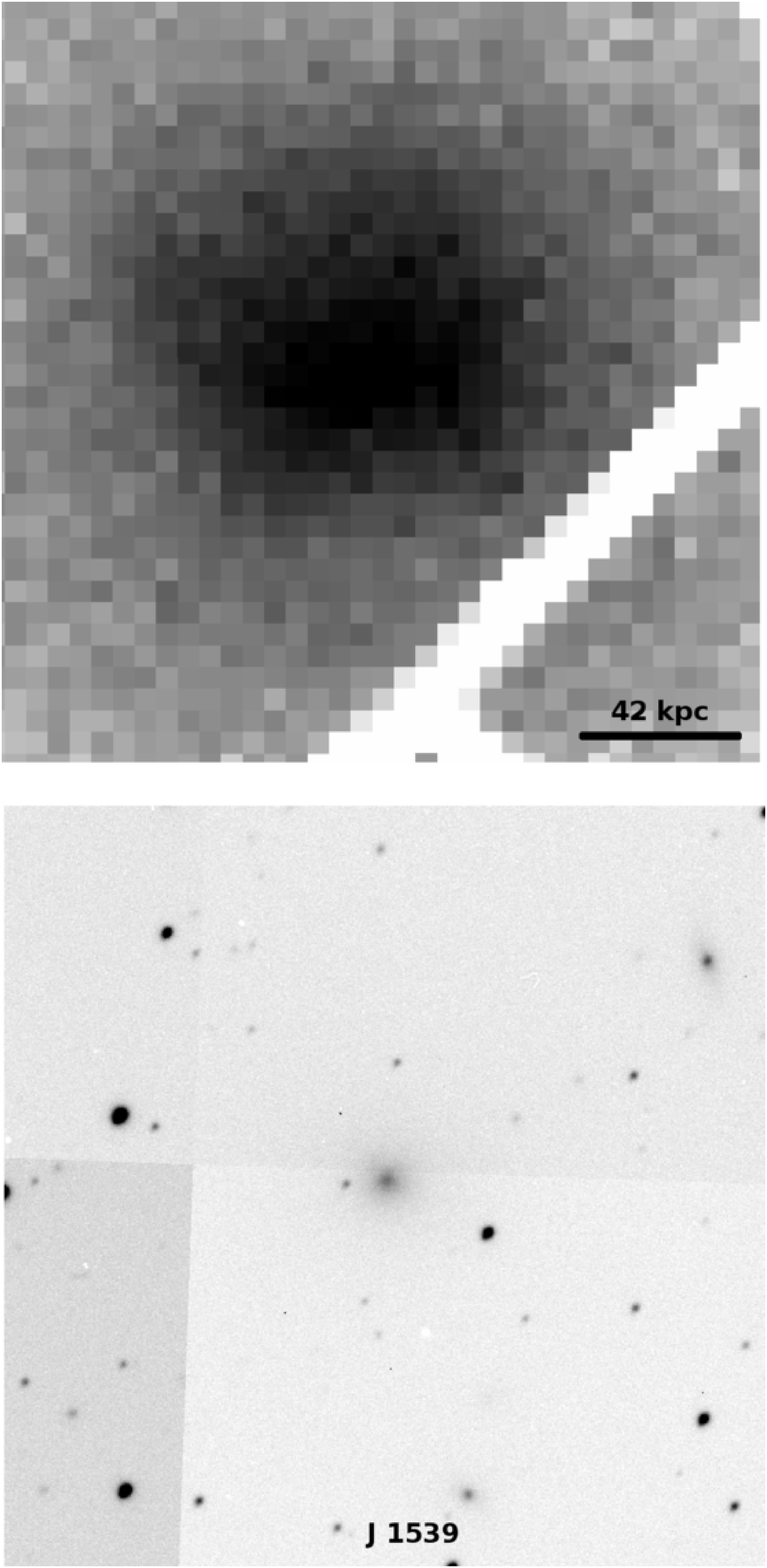}
 \includegraphics[width=0.3\linewidth]{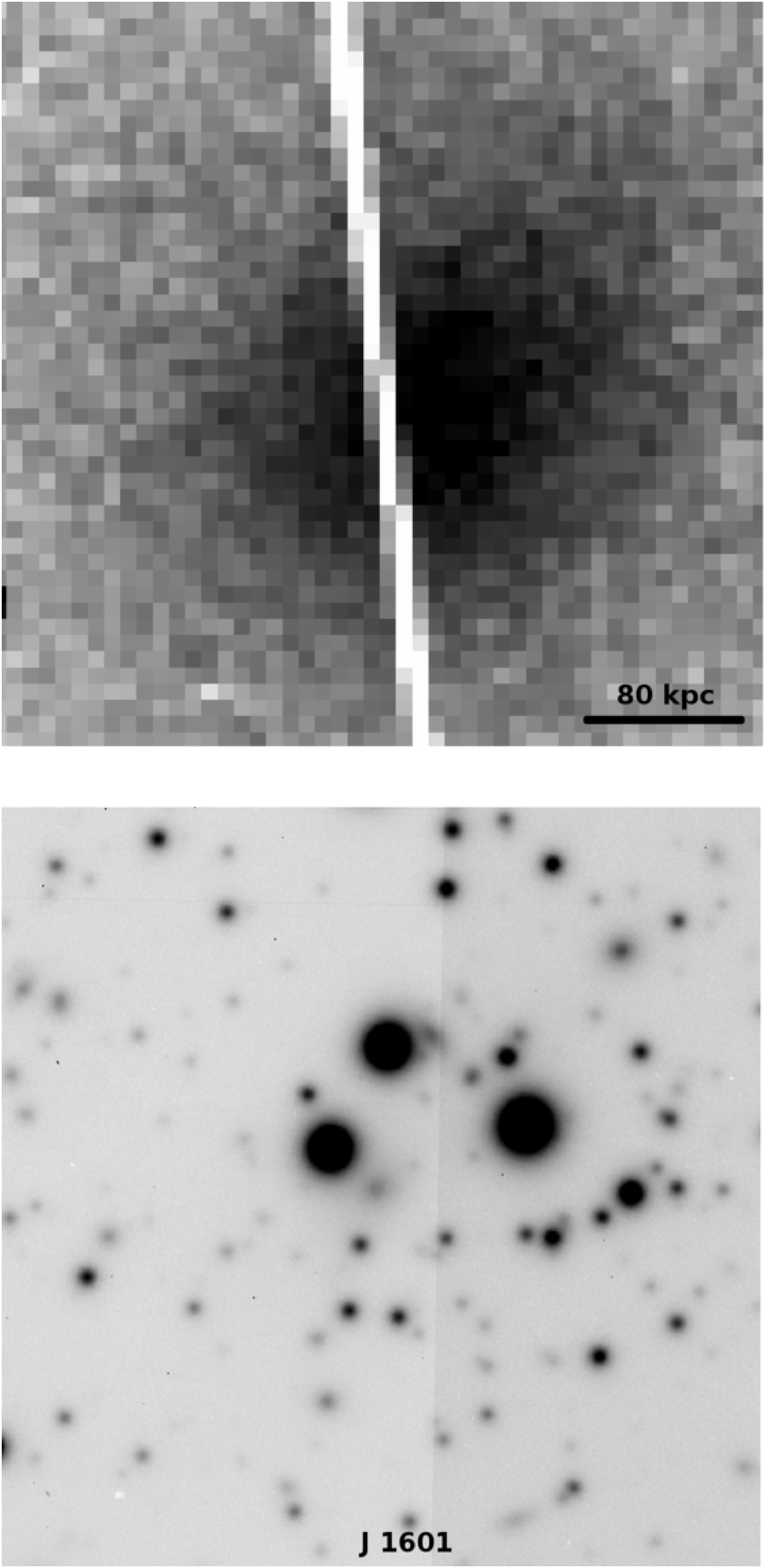}
 \includegraphics[width=0.3\linewidth]{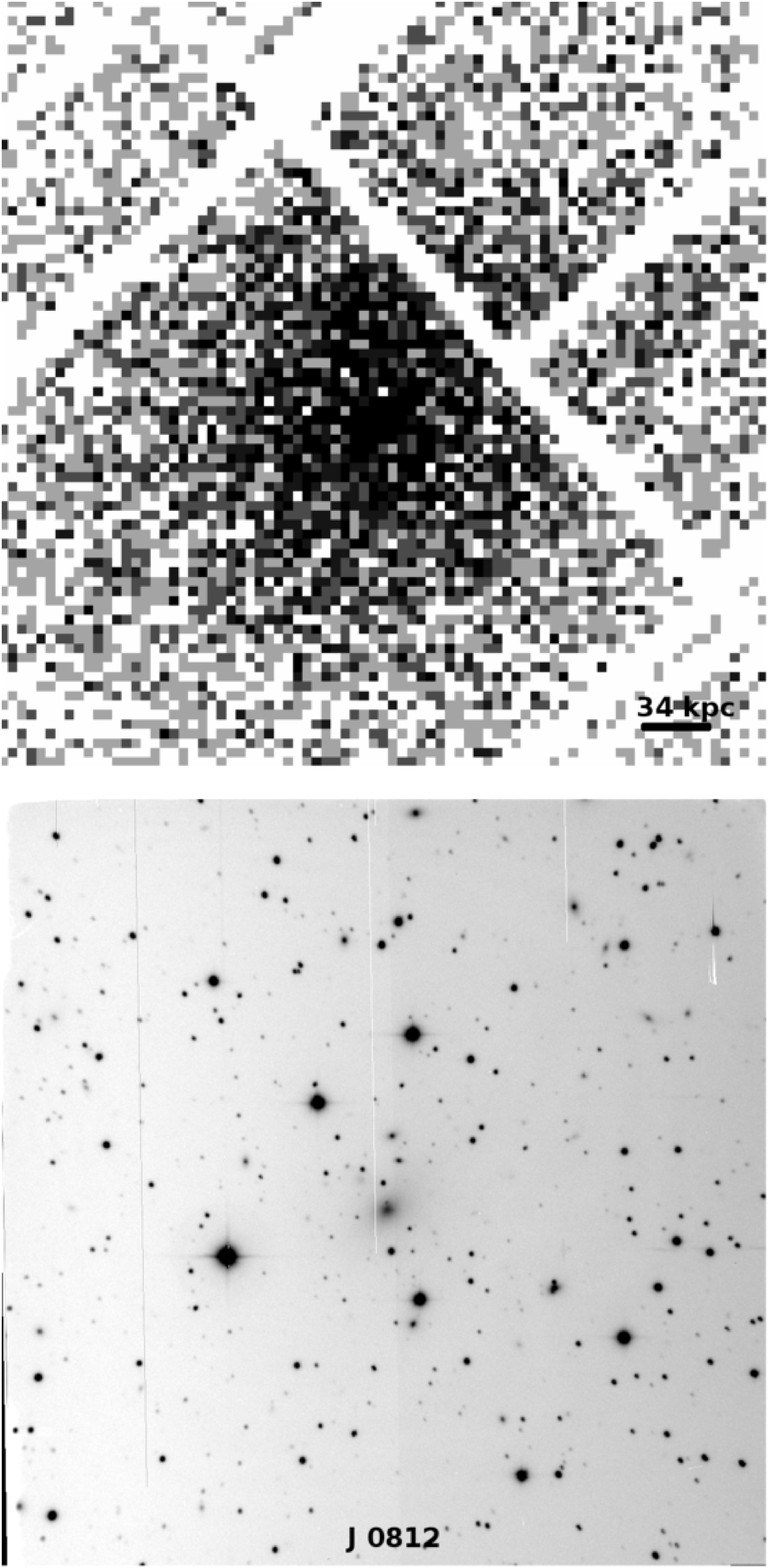}
}
\caption{X-ray (top) and optical (bottom) images of the studied clusters. Top and bottom image are at the same scale and the length of the scale bars is \asec[30]. \src{1101} and \src{0301} both have a clear cD galaxy in the centre. In \src{0501} there are three bright foreground stars in the field of view, which makes any optical central structure in this cluster hard to see.}\label{fig:optX}
\end{figure*}
\begin{figure}[t]
 \centerline{\includegraphics[angle=-90,width=0.9\linewidth]{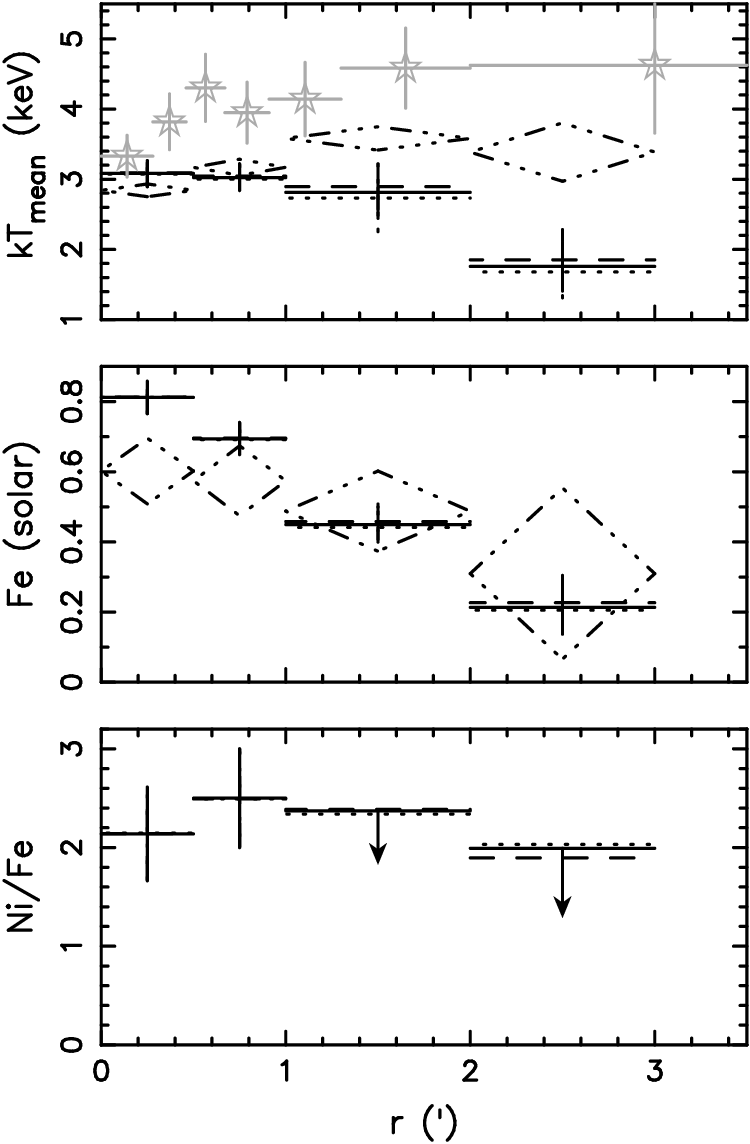}}
\caption{Radial profiles of temperature, Fe abundance and Ni/Fe abundance ratio for \src{1101}. The solid crosses represent the values for subtracting the background using the vignetting factor listed in Table \ref{tab:results}. The dashed crosses represent the result when subtracting the background using the vignetting factor minus its statistical uncertainty, the dotted crosses represent the case when the vignetting plus its statistical uncertainty is used for subtraction. The dash-dotted diamond shapes represent the values fitted to the Chandra ACIS data in the same spatial region. The grey stars represent the values fitted by \citet{CAV09} using the Galactic \ion{H}{i} value for the hydrogen column density. The arrows in the bottom panel represent 2$\sigma$ (95.4\% confidence level) upper limits. }\label{fig:prop_1101}
\end{figure}

A summary of the parameters fitted to the data from the \amin[3] around the centre of the three clusters (which we will hereafter refer to as the ``integrated'' spectrum) is given in Table \ref{tab:results}. The nominal statistical uncertainties of the fits are shown, as well as (in brackets) the uncertainty associated with the statistical uncertainty of the subtracted background. In Fig. \ref{fig:optX}, optical and X-ray images of the sources are shown. We also fitted the spectra of different annuli employing the same model as for the integrated spectrum (Figs. \ref{fig:prop_1101} and \ref{fig:prop_0501}). In the following sections we describe the results for each source.
\begin{table}[t]
\begin{center}
\caption{Summary of the fits to the \Chandra ACIS spectra of \src{1101}. }\label{tab:chandra}
\begin{tabular}{ccc}
region\tnote{a} & \textit{kT} & Fe abundance \\
& (keV) & (proto-solar) \\
\hline
\hline
1 & 3.02 $\pm$ 0.13 & 0.9 $\pm$ 0.3\\
2 & 2.7 $\pm$ 0.2 & 0.41 $\pm$ 0.14\\
res\tnote{b} & 3.28 $\pm$ 0.14 & 0.52 $\pm$ 0.11 \\
\hline
\end{tabular}
\end{center}
\tnotetxt{a}{The regions are shown in Fig. \ref{fig:chandra}.}
\tnotetxt{b}{For this spectrum, regions ``1'' and ``2'' are excluded}
\end{table}
\begin{table*}[ht]
\begin{center}
\caption{Best-fit parameters for spectra of the \amin[3] around the centre of the galaxy clusters. } 
\label{tab:results}
\begin{tabular}{lllll}
\hline 
 Parameter  &  \src{1101} (\XMM)  \tnote{a}$^{,}$\tnote{b}  & \src{1101} (\Chandra)\tnote{a} &  \src{0501}\tnote{a}  &  \src{0301} \\
 \hline 
 \hline 
 Background scaling factor  &  0.99 $\pm$ 0.02  &   0.976 $\pm$ 0.013   &  --  \\
\hline
 Hydrogen column density (\unit[$\mathrm{10^{24}}$]{m^{-2}})  &  20.0 $\pm$ 0.4 ($\pm$ 0.2)  & 17.2 $\pm$ 1.3 &   6.6 $\pm$  0.4 ($\pm$ 0.3)  &  29 $\pm$ 2  \\
  Hydrogen \unit[21]{cm} column density\tnote{d} (\unit[$\mathrm{10^{24}}$]{m^{-2}}) & 8.38 & 8.38 & 6.25 & 22.1 \\
Emission measure\tnote{c} (\unit[$10^{73}$]{m^{-3}}) & 2.57  $\pm$  0.04 ($\pm$ 0.01) & 2.44 $\pm$ 0.07 & 9.2  $\pm$  0.2 ($\pm$ 0.2) & 2.80 $\pm$ 0.16 \\
 Temperature (keV)\tnote{e}  &  3.03  $\pm$  0.15 ($\pm$ 0.08)  & 3.50 $\pm$ 0.11 &  8.24 $\pm$ 0.58  ($\pm$ 0.97) &   3.2 $\pm$ 0.2  \\
 $1/\alpha$\tnote{f}  &  0.48 $\pm$ 0.09 ($\pm$ 0.03) & 0.48 (fixed) & --   &  --  \\
  Fe abundance  &  0.61  $\pm$ 0.03  ($\pm$ 0.02) & 0.57 $\pm$ 0.07 &   0.58 $\pm$  0.10   ($\pm$ 0.06)  &    0.37 $\pm$ 0.10  \\
  Si/Fe abundance   &  0.65  $\pm$  0.12 ($\pm$ 0.01)  & -- &   --   &   --  \\
  Ni/Fe abundance  &  2.81   $\pm$  0.56 ($\pm$ 0.02)  & -- &   --  &    --  \\
 Flux (\unit[0.1 -- 2.4]{keV}) [\unit[$\mathrm{10^{-14}}$]{W\,m^{-2}}]  &  1.03 $\pm$ 0.02  ($\pm$ 0.01)  & 1.03 $\pm$ 0.03 &  0.99   $\pm$  0.02 ($\pm$ 0.02)  &  0.66  $\pm$ 0.04  \\
 Flux (\unit[2 -- 10]{keV}) [\unit[$\mathrm{10^{-14}}$]{W\,m^{-2}}]  & 1.11  $\pm$  0.02 ($\pm$ 0.04)  & 1.23 $\pm$ 0.03 & 1.70 $\pm$  0.04 ($\pm$ 0.13)  &   0.88 $\pm$ 0.05 \\
 C-stat / d.o.f.   &   270/229   & 389/295 &  252/224   & --\\
$\chi^{2}$ / d.o.f.    & -- & -- & -- &  189/134  \\
\hline 
\end{tabular}
\end{center}
     \tnotetxt{a}{The first quoted uncertainty is the statistical uncertainty on the value. The uncertainty between parentheses is due to the uncertainty in background estimation (Sect. \ref{subsec:flares}).}
        \tnotetxt{b}{For this source, we fitted a ``\textit{wdem}'' model.}
        \tnotetxt{c}{The emission measure is defined as $Y=\int n_{\mathrm{e}} n_{\mathrm{H}} \mathrm{d}V$, see also Sect. \ref{sec:models}}
        \tnotetxt{d}{Deduced from the \unit[21]{cm} map by \citet{DIC90}.}
         \tnotetxt{e}{For \textit{wdem}: ``mean temperature'', see Eq. \eqref{eq:wdem_mean}}  
         \tnotetxt{f}{As defined in Eq. \eqref{eq:wdem_dydt}.}
\end{table*}
\subsection{\src{1101}} \label{sec:res_1539}
The integrated spectrum of our brightest source, \src{1101}, is shown in Fig. \ref{fig:spec_1101}. As shown in the plot, the background is high compared to the source (see also Sect. \ref{subsec:flares}). The effect is noticeable for the part of the spectrum above \kev[2.5]. We fitted a \textit{wdem} model to the background-subtracted spectrum. The mean temperature of the source is \kev[3.03$\pm$0.15].  
\par Visual inspection of the \Chandra image of \src{1101} (see Fig. \ref{fig:chandra}) shows complexity in the cluster morphology: a compact bubble-shaped region and an extended structure in the upper part of it stand out from the weaker diffuse emission of the cluster. The three regions indicated in the image yield the parameters cited in Table \ref{tab:chandra}. Due to the relatively low statistics, we fitted a single-temperature model to these spectra. The region labelled ``2'' has a temperature of \kev[2.7$\pm$ 0.2], which is slightly below the temperature of the surrounding gas of \kev[3.28 $\pm$ 0.14]. This is consistent with the values found in the \XMM analysis. Fits to the \amin[0-3] region of the \Chandra observation yield a hydrogen column density of \unit[(17.2 $\pm$ 1.3)$\times\mathrm{10^{24}}$]{m^{-2}}. This is consistent with the best-fit column density derived for the \XMM data, \unit[(20.0 $\pm$ 0.4 ($\pm$ 0.2))$\times\mathrm{10^{24}}$]{m^{-2}}. Here, the uncertainty between parentheses is due to the uncertainty in the background determination by comparing the values of maximum and minimum background scaling factors.
\subsection{\src{0501}}
\begin{figure}[t]
 \centerline{\includegraphics[angle=-90,width=0.9\linewidth]{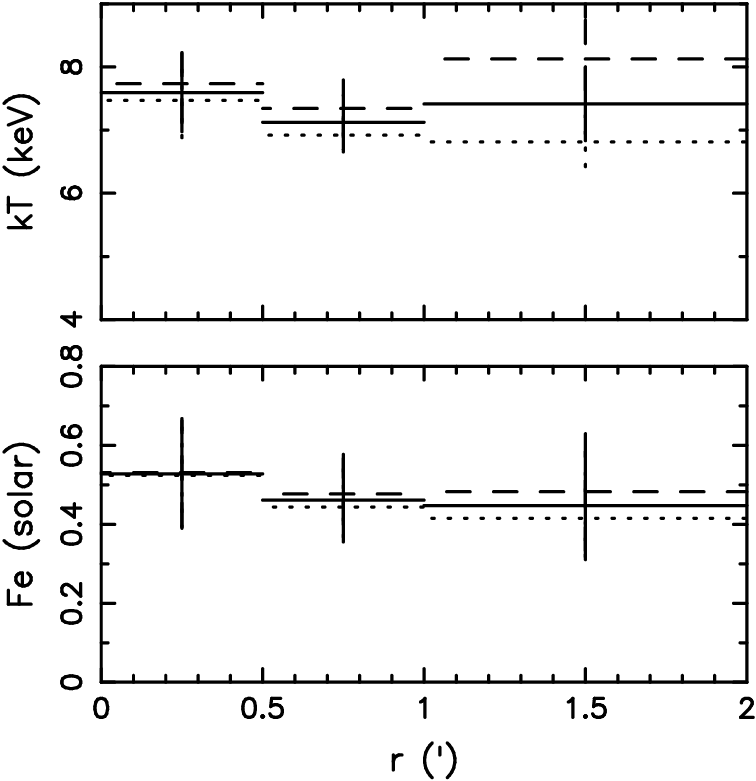}}
\caption{Radial profiles of temperature and Fe abundance for \src{0501}. The solid crosses represent the values for subtracting the background using the vignetting factor cited in Table \ref{tab:results}. The dashed crosses represent the result when subtracting the background using the vignetting factor minus its statistical uncertainty, the dotted crosses represent the case when the vignetting plus its statistical uncertainty is used for subtraction.
}\label{fig:prop_0501}
\end{figure}
%
The temperature fitted for the integrated spectrum of this source is \kev[8.24$\pm$0.58($\pm$0.97)] (see Table \ref{tab:results}). The fit is acceptable (see Table \ref{tab:results}). We did not find significant multi-temperature structure for this source. 
Figure \ref{fig:prop_0501} shows that there is a flat radial temperature and iron abundance distribution. Silicon and nickel abundances were not measured in any significant way.
\subsection{\src{0301}}
This is the only source for which the background during the observation was not high. Because we multiplied the counts in the image with the theoretical vignetting weighting, the uncertainties are not the square root of the measured counts anymore. In this case, the Poisson statistic is not valid for the observation, and we used a fit based on the $\chi^{2}$. The temperature of the integrated spectral region is measured to be \kev[3.2$\pm$0.2]. The Fe abundance is 0.37$\pm$0.10 solar. No other abundances could be significantly determined.
\section{Discussion and Conclusions}\label{sec:disc}
We reported the spectral properties of three moderately bright clusters, which were selected from \ac{WFC} data. Radial temperature and abundance profiles of the the cores of \src{1101} and \src{0501} were obtained. In addition, the temperature and iron abundance in the highly absorbed cluster \src{0301} are measured. We were able to derive accurate temperatures and abundances, despite the high soft-proton background in two of our cluster observations. In Sect. \ref{subsec:flares}, we described a novel method to estimate the vignetting factor of soft  protons for \pn. The systematic uncertainties introduced by this new method are mostly similar to or smaller than the statistical uncertainties found in our spectral fits. From the plots in Figs. \ref{fig:prop_1101} and \ref{fig:prop_0501}, it is clear that the effect of the background subtraction becomes more important further away from the centre of the field of view due to the distribution of the source brightness. Especially in the 2--3\amin region of the upper panel of Fig. \ref{fig:prop_1101} it is clear that this effect is non-negligible when measuring temperatures.
\par The difference we fitted between the \kev[0.1--2.4] fluxes we found as well as the fluxes from both the ROSAT all sky survey \citep{VOG99} and the REFLEX catalogue \citep{BOH04} are typically 20--30\%. A comparison of the fluxes with catalogue values is however not straightforward. The measured flux depends on the model used and the extraction region (in our case \amin[3]). Furthermore, the hydrogen column density used in these catalogues is based on the \ion{H}{i} data from \citet{DIC90}. The value we used is found by fitting the spectra. We found a higher column density in two of our cases. For \src{1101}, the higher column density is also found with the \Chandra data.
\subsection{\src{1101}}
\src{1101} is the brightest source of our cluster sample.
We found a hydrogen column density that is a factor 2.4 higher than the value from \citet{DIC90}. This value is consistent between \XMM and \Chandra.
The mean temperature of the integrated spectral region for this cluster is about \kev[3], which indicates that \src{1101} is a relatively cool cluster. This source also had a significant multi-temperature structure when we fitted the spectrum with a differential emission measure (DEM) model.
The strong radial iron gradient in \src{1101} is similar to abundance gradients found in cool-core clusters \citep{DEG04,TAM04}. In the \Chandra data, no clear trend is visible, which is due to the large error bars of the outer data points.
\par The \Chandra image shows structure in the core (see Fig. \ref{fig:chandra}). The central elliptic region (labelled ``1'') is cooler than its surroundings, and the top region (labelled ``2'') is cooler than the rest of the gas, which is at comparable distances from the core (labelled ``res.''). The Fe abundance is consistent with the value derived from the XMM-Newton observation. There is a slight hint for a higher Fe abundance in the region ``1''. The morphology looks similar to other examples of cool-core clusters where bubbles were observed, for example in \object{Abell 2052} \citep{BLA01} and \object{Hydra A} \citep{MCN00}.
The bubbles in clusters are thought to be caused by \ac{AGN} outbursts in the central galaxy \citep[for a review on this topic see][]{MCN07}. These outbursts are detected in the radio domain and create cavities in the X-ray emission of the gas. These AGN also seem to provide the energy needed to keep the gas from cooling below $\sim$ \kev[0.5]. Gas at these temperatures is not observed in clusters of galaxies \citep{TAM01,KAA01,PET01,PET03,KAA04}, therefore a reheating process of the gas is needed. 
\par In the top panel of Fig. \ref{fig:prop_1101} we present the temperature profiles derived from our fits, and compared them to the profile derived by \citet{CAV09}. Both our radial \XMM and \Chandra temperature profiles are significantly different from their profile because they keep the hydrogen column density fixed to the \ion{H}{i} value. We let the hydrogen column density free in our fits, and found a value which is 2.4 times higher in both the \Chandra and the \XMM data sets.
\begin{figure}[ht]
 \includegraphics[width=0.35\textwidth, angle=-90]{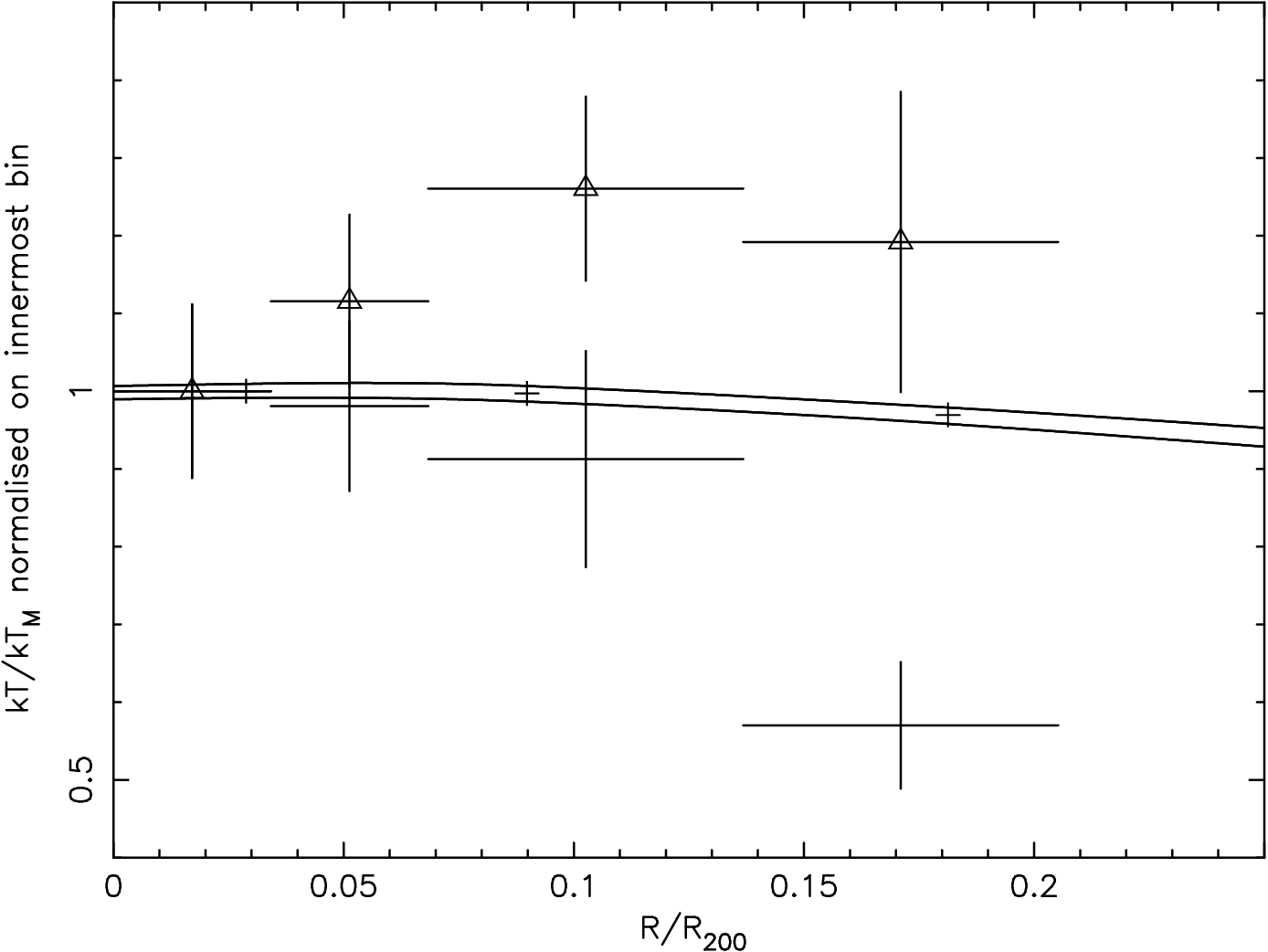}
\caption{Comparison of the \XMM and \Chandra radial temperature profiles found with the plot of the uncertain cool-core cluster sample from \citet{LEC08a}. The small crosses represent the data points from \citet{LEC08a}. The lines represent the 1 $\sigma$ upper and lower limit on the sample. The large crosses are the data points of our \XMM observation. The crosses with triangles are the points obtained by fitting the \Chandra data. Because of the difference in definition of the mean temperature, the temperature values are all scaled to the innermost bin. The $R_{200}$ is calculated using the relation derived by \citet{ARN05}}\label{fig:UCC}
\end{figure}
\par The radial temperature profile derived from the \XMM data shows a strong decline in the \amin[1--3] region. The \Chandra data however show a much weaker decline. We compared the radial temperature profiles with a profile, based on average profiles compiled by \citet{LEC08a}.
They show profiles for the cooling-core, unidentified cooling core and non-cooling core clusters. This classification is based on the radial behaviour of temperature in the cluster core. Since we did not measure a significant temperature decrease with \XMM in this region, we compare our results with the average of the unidentified cooling core clusters. 
\par The temperature $kT_{mean}$ in \citet{LEC08a} is defined by fitting a single temperature model to the region between $0.1 R_{180}$ and $0.6 R_{180}$ \footnote{In \citet{LEC08a}, $R_{180}$ is used. Based on their definition of that radius and the definition of the $R_{200}$ from \citet{ARN05} we find that $R_{200}=1.012 R_{180}$.}. However, the annuli we fitted extend only to $0.2 R_{200}$. Therefore we could not use this range for our mean temperature. To compare the two data sets with the trend, we scaled the relative temperature to the value of the innermost bin. This results in the plot shown in Fig. \ref{fig:UCC}. 
\par Since the shape of the \Chandra and \XMM temperature profiles are different, the scaling on the innermost bin results in more significant deviations between the profiles at larger radii (see Fig. \ref{fig:UCC}). The absolute temperatures in the three inner bins lie within $2.1\sigma$. Furthermore, the scaled \XMM data point at $0.17 R_{200}$ is $4.7\sigma$ below the average profile of \citet{LEC08a}.
\par The difference in profile shapes between \XMM, \Chandra and the average profile can be explained in multiple ways. The cooler temperature of the innermost bin as measured by \Chandra may be partly due to \ac{PSF} effects. The larger PSF of \XMM may cause photons from the second radial bin to scatter into the central bin and subsequently raise the derived temperature. In addition, the subtracted soft-proton background may not exactly represent the true soft-proton spectrum as measured in the annuli. This is because the soft-proton spectrum in the inner annuli might have a slightly different slope compared to the soft-proton spectrum in the outermost annulus on which the subtracted background is based. The slope of the subtracted background can have a significant influence on the measured temperature, especially in the outer annuli where the background is high with respect to the source spectra.
\subsection{\src{0501}}
The hydrogen column density that we fitted for \src{0501} is consistent with the value based on the \ion{H}{i} maps of \citet{DIC90}. We derived a temperature for the integrated spectrum, which is rather high for clusters of galaxies. Clusters with high temperatures have less prominent spectral lines, and the fit of the temperature is mostly dependent on the continuum. Therefore, the uncertainty on the background scaling factor introduces a large systematic uncertainty on the temperature. 
\par The flat temperature and iron abundance profiles suggest however that \src{0501} is a non-cooling core cluster. This means either that the core of the cluster has not cooled yet or was recently reheated. The rate at which the cluster core cools can be estimated by the cooling time. This is the typical time scale during which the core should have radiated all its energy away. \citet{SAR88} gives for the cooling time of a cluster of galaxies
\begin{equation}
t_{\mathrm{cool}}=8.5\times10^{10} \left ( \frac{n_{\mathrm{p}}}{10^{3} \mathrm{m^{-3}}}\right )^{-1} \left( \frac{T_{\mathrm{g}}}{10^{8}\mathrm{K}} \right)^{1/2} \mathrm{yr} \text{.}
\end{equation}
In this equation, $T_{\mathrm{g}}$ is the gas temperature and $n_{\mathrm{p}}$ represents the proton density. The volume is calculated with the Cosmology Calculator by \citet{WRI06}. We used the emission measure that we found for the innermost \asec[30] of the cluster core (\unit[\pten{(1.98 \pm 0.08)}{73}]{m^{-3}}) and the temperature of the same region (7.4$\pm$0.7 \kev), assuming that this region is a sphere with constant density. We took the electron density to be a factor 1.2 higher than the proton density. The cooling time we compute is 4.9$\pm$0.2 Gyr, which is about 1/2 of the Hubble time at the redshift of \src{0501}, which indicates that the cluster has had enough time to cool. 
\par A merger at an early stage of the clusters life could cause shocks, which would reheat the gas. This could explain the absence of a cool core  when the cooling time is more than $\sim 0.1$ Hubble times \citep[e.g.][]{BUR97,GOM02}. The image of this cluster made using \XMM (top-middle panel of Fig. \ref{fig:optX}) and the intensity contours of the ROSAT \ac{HRI} (Fig. \ref{fig:0501_rosat}) hint towards an elongation of the X-ray cluster emission. This hint is however not statistically strong enough to pinpoint a merger. The bottom-middle panel of Fig. \ref{fig:optX} shows three very bright foreground stars, which make it hard to state whether there is one central galaxy or if the core consists of multiple galaxies, which could indicate a merger history.
\subsection{\src{0301}}
The \kev[0.1--2.4] flux we derived for this source is 60\% lower than the ROSAT value of \citet{VOG99}. The integrated temperature is relatively low and the iron abundance is slightly lower than the average of larger cluster samples \citep[e.g.][]{TAM04,DEG04,PLA06}. The hydrogen column density we fitted is high, 36\% higher than the value of \citet{DIC90}. 
Although this is the only observation for which the soft-proton background is very low, it is also the dimmest of the three sources. Therefore, the statistical quality of this observation is limited.
\begin{acknowledgements}
We thank the anonymous referee for constructive comments. We also thank Eva Ratti for helping us identify the stars in \src{0501}. Based on observations obtained with XMM-Newton, an ESA science mission with instruments and contributions directly funded by ESA Member States and NASA.The Netherlands Institute for Space Research (SRON) is supported financially by NWO, the Netherlands Organisation for Scientific Research. 
\end{acknowledgements}
\bibliographystyle{aa} 
\bibliography{biblio} 
\end{document}